\documentclass[aps,pra,reprint,twocolumn,showpacs,superscriptaddress,floatfix]{revtex4-1}
\usepackage{amsmath}
\usepackage{amssymb,amsfonts}
\usepackage[dvips]{graphicx}
\usepackage[small]{subfigure}
\usepackage[usenames,dvipsnames]{xcolor}

\newcommand{\fig}[1]{Figure~\ref{#1}}
\newcommand{\be}[1]{\begin{equation}\label{#1}}
\newcommand{\ee}{\end{equation}}

\begin{document}
\title{Multiple electron trapping in the fragmentation of strongly driven molecules}
\author{A. Emmanouilidou}
\affiliation{Department of Physics and Astronomy, University College London, Gower Street, London WC1E 6BT, United Kingdom} 
\affiliation{Chemistry Department, University of Massachusetts at Amherst, Amherst, Massachusetts 01003, USA}
\author{C. Lazarou}
\affiliation{Department of Physics and Astronomy, University College London, Gower Street, London WC1E 6BT, United Kingdom} 
\begin{abstract}
We present a theoretical quasiclassical study of the formation, during Coulomb explosion, of two highly excited neutral H atoms (double H$^{*}$) of strongly driven H$_2$. In this process, after the laser field is turned off each electron occupies a Rydberg state of an H atom. We show that two-electron effects are important in order to correctly account for double H$^{*}$ formation. We find that the route to forming two H$^{*}$ atoms is similar to pathway B that was identified in Phys. Rev. A {\bf 85} 011402 (R) as one of the two routes leading to single H$^{*}$ formation. However, instead of one ionization step being ``frustrated"  as is the case for pathway B, both ionization steps are ``frustrated" in double H$^{*}$ formation. Moreover, we compute the screened nuclear charge that drives the explosion of the nuclei during double H$^{*}$ formation.

\end{abstract}
\maketitle
Many interesting physical phenomena arise during fragmentation of molecules driven by intense infrared laser fields. 
Examples of such phenomena are bond-softening and above-threshold ionization \cite{Suzor1990,Zavriyev1990}, molecular 
non-sequential double-ionization \cite{Niikura2002,Staudte2002,Sakai2003,Alnaser2003} and 
enhanced ionization \cite{Niikura2002,Zuo1995,Seideman1995,Villeneuve1996,Dehganian2010}.
Another interesting phenomenon, the formation of single highly excited 
neutral fragments, has been recently observed in strongly driven H$_2$ \cite{Manschwetus2009} and other molecules \cite{Nubbemeyer2009,Manschwetus2010,Ulrich2010,McKenna2011}. 
 The formation of excited fragments has been attributed to  ``frustrated" tunnel-ionization \cite{Nubbemeyer2008}. In a very recent theoretical study of strongly driven H$_{2}$ \cite{Emmanouilidou2012}, two main routes were identified leading to single H$^{*}$ formation. These two routes differ on whether the  first (pathway B) or second (pathway A) ionization step is ``frustrated" .
   
In this work, using the theoretical model developed in \cite{Emmanouilidou2012}, we study the formation of two highly excited H$^*$ atoms during the breakup of 
H$_2$ by intense infrared laser fields. Multiple electron capture in Rydberg states was recently observed in strongly-driven Ar$_2$ dimers \cite{Dorner2011}; no break-up channel with two neutrals as end fragments was reported in this latter study.
 We show that, as  for pathway B of single H$^{*}$ formation, two-electron effects are important in the pathway leading to double H$^{*}$ formation. Moreover, we quantify how the two highly excited electrons in double H$^{*}$ formation screen the nuclear charge and compute the  effective charge that drives the Coulomb explosion of the nuclei.

Our quasiclassical model was discussed in detail in reference  \cite{Emmanouilidou2012}. 
It fully accounts for both the electronic and nuclear motion in order to accurately describe, among other processes, the formation of one and two H$^{*}$ atoms. Treating both electronic and nuclear motion in theoretical studies of strongly-driven molecules is a challenging task. This is the reason why, with few exceptions  \cite{Leth2009,Martin2009}, in most studies either the nuclei are fixed \cite{Alnaser2003,Manohar2006} and only the electronic motion is considered or the electronic continuum is ignored and only the nuclear motion is studied \cite{McKenna2008}. 

For completeness, we briefly describe our quasiclassical model. First, we set up the initial electronic phase space distribution. We consider parallel alignment between the molecular axis and the laser electric field (along the z axis) to complement our studies of single H$^{*}$ formation in \cite{Emmanouilidou2012}.
The field is taken to be $\mathrm{E(t)=E_{0}(t)\cos({\omega t})}$ at 800 nm  corresponding to
$\mathrm{\omega= 0.057}$ a.u. (a.u. - atomic units).
 In our simulation, the pulse envelope $\mathrm{E_{0}(t)}$ is defined as $\mathrm{E_{0}(t) = E_{0}}$
for $\mathrm{0 < t < 10T}$ and $\mathrm{E_{0}(t) = E_{0}\cos^2(\omega(t-10T )/8)}$ for
$\mathrm{10T < t < 12T}$ with T the period of the field.  We start the time propagation at $\mathrm{\omega t_{0}=\phi_{0}}$  where the phase of the laser field $\mathrm{\phi_{0}}$ is chosen randomly, see \cite{Chen2000,Brabec1996,Emmanouilidou2009}.
 If the instantaneous field strength at phase $\mathrm{\phi_{0}}$ is smaller than the threshold field strength for over-the-barrier ionization, we assume one electron (electron 1) tunnel-ionizes, i.e., tunnels through the field-lowered Coulomb potential to the continuum  whereas the initially bound electron (electron 2) is described by a one-electron microcanonical distribution.  If the instantaneous field strength at phase $\mathrm{\phi_{0}}$ allows for  over-the-barrier ionization we use a double electron microcanonical distribution (see \cite{Emmanouilidou2009}). For both intensity regimes we use the tunneling rate provided by the semiclassical formula in reference  \cite{Li2007} with field strength equal to the instantaneous one at $\mathrm{\phi_{0}}$.  We use 0.57 a.u. (1.28 a.u.) as the first (second) ionization potentials.
 Second, for the initial phase space distribution of the nuclei we use the Wigner function of the ground state (energy 0.01 a.u.) of the Morse potential \cite{Frank2000}; the parameters in the latter potential best describe the ground vibrational state of H$_2$, for details see \cite{Emmanouilidou2012}.

Third, we transform to a new system of  ``regularized" coordinates \cite{Kustaanheimo1965, Heggie1974}. This
transformation explicitly eliminates the Coulomb singularity
\cite{Emmanouilidou2009}. We then
propagate the full four body Hamiltonian
using the classical trajectory Monte Carlo method \cite{Abrines1966}.  During the time propagation we allow the initially 
bound electron to tunnel at the classical turning points along the molecular axis using the WKB approximation \cite{Cohen2001, Merzbacher}. Finally 
we select only those trajectories that during the breakup of H$_2$ result in the H$^{*}$+H$^{*}$ channel (where $*$ denotes an
electron in a $\mathrm{n>1}$ quantum state). To identify the electrons captured in a Rydberg n quantum state of $\mathrm{H^{*}}$
 we follow the method described in  reference  \cite{Becker1984}.
 \begin{figure}
\begin{center}
\includegraphics[width=0.9\columnwidth]{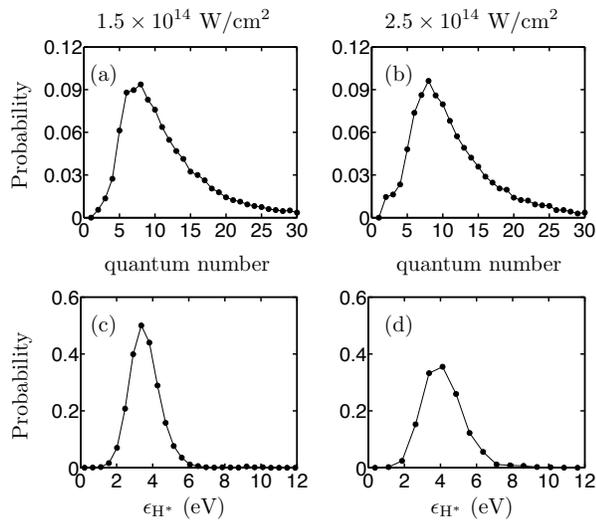}
\caption{Top row: The distribution of the quantum number $\mathrm{n}$ of the final Rydberg state occupied by either electron in the double H$^{*}$ formation channel. Bottom row:  The final energy distribution of either of the H$^*$ fragments. The left column is for an intensity $1.5\times10^{14}$ W$/$cm$^2$ and 
the right one is for $2.5\times10^{14}$ W$/$cm$^2$.}\label{fig1}
\label{figure1}
\end{center}
\end{figure}

In order to study the intensity dependence of double H$^*$ formation we have considered an intensity of $1.5\times10^{14}$ W$/$cm$^2$ in the tunneling regime and an intensity of $2.5\times10^{14}$ W$/$cm$^2$  in the over-the-barrier regime; however, for the latter intensity, most of the
trajectories are initiated with the tunneling model. We compute the final $\mathrm{n}$ quantum number  [see \fig{figure1}(a) and (b)] for either H$^{*}$ fragment for both intensities; at least 20000 double H$^{*}$ events are considered.
 The final n quantum number distribution  peaks at $\mathrm{n=8}$ for both intensities resembling the n distribution for singly excited neutrals formed  either in strongly driven H$_2$ \cite{Emmanouilidou2012} or strongly driven atoms \cite{Nubbemeyer2008}.
Moreover, the final energy distribution of either H$^{*}$ fragment in the H$^*+$H$^*$ channel [see \fig{figure1}(c) and (d)] resembles that for the H$^+$ or H$^*$ fragments in the H$^*+$H$^++e^{-}$ channel \cite{Emmanouilidou2012}. For the lower intensity [\fig{figure1}(c)] the energy distribution peaks at approximately 3.5 eV.  For the higher intensity [\fig{figure1}(d)] the peak shifts to higher energies  since the nuclei Coulomb explode at smaller internuclear distances.

 \begin{figure}
\begin{center}
\includegraphics[width=0.9\columnwidth]{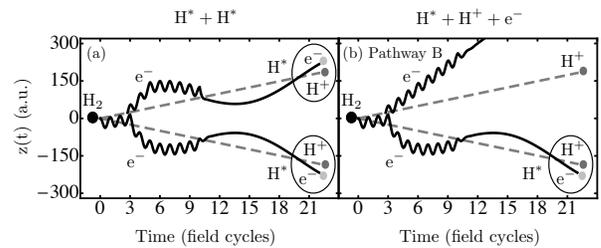}
\caption{(Colour online) Schematic illustration of the route leading to double H$^*$ formation (a) and that of pathway B of single H$^*$ formation (b). Shown is the time-dependent position along the laser field for electrons (black lines) and ions (grey broken lines).}
\label{route}
\end{center}
\end{figure}

In \fig{route}(a) we show the pathway leading to  double $\mathrm{H^{*}}$ formation where electron 1 tunnel-ionizes very quickly, quivering in the laser field, while electron 2 tunnel-ionizes after a few periods of the laser field. However, when the field is turned off, both electrons 1 and 2 do not have enough drift energy to escape and occupy instead a Rydberg state of an H atom. Similarly,  in pathway B for single H$^{*}$ 
formation \cite{Emmanouilidou2012}, see \fig{route}(b), electron 1 tunnel-ionizes very quickly, quivering in the field, while electron 2 tunnel-ionizes and escapes after a few periods of the laser field; however, when the field is turned off, electron 1 does not have enough drift energy to escape and occupies a Rydberg state of an H atom.  Hence, while in the pathway leading to double H$^*$ formation both ionization steps are ``frustrated" in pathway B only the first one is.  In what follows, we show that in all other respects the break-up channel leading to two H$^{*}$ atoms and pathway B of single  H$^{*}$ formation are similar.   
   
\begin{figure}
\begin{center}
\includegraphics[width=0.9\columnwidth]{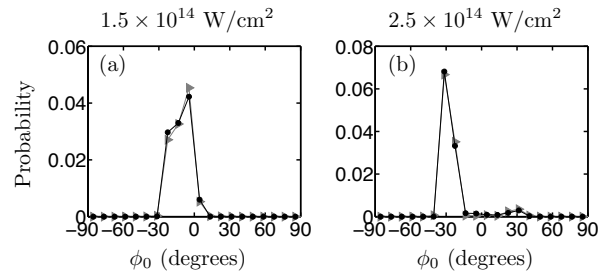}
\caption{The distribution of the field phase $\mathrm{\phi_0}$ at the time when electron 1 tunnel-ionizes in the initial state for the double H$^{*}$ formation channel $(\bullet)$, and for pathway B of the single H$^{*}$ formation  channel $({\color{Gray}{\blacktriangleright}})$. Figure (a) is for an intensity of $1.5\times10^{14}$W$/$cm$^2$ and (b) for $2.5\times10^{14}$W$/$cm$^2$.}\label{fig4}
\label{laserphase}
\end{center}
\end{figure} 
        
We explore the route to double H$^{*}$ formation by, first, plotting in \fig{laserphase} the distribution of the field phase, $\mathrm{\phi_0}$, when electron 1 tunnel-ionizes in the initial state.  Two H$^{*}$ atoms form when $\phi_0$ is around 0$^{\circ}$ for the lower intensity [\fig{laserphase}(a)] and around -30$^{\circ}$ for the higher intensity [\fig{laserphase}(b)].  The reason double H$^{*}$ formation shifts from small values (extrema of the field) to larger values with increasing intensity  is the onset of saturated ionization of the neutral molecule \cite{Emmanouilidou2009}. Comparing the distributions of the field phase $\mathrm{\phi_0}$ for pathway B of single H$^{*}$ formation and for double H$^*$ formation  in \fig{laserphase}(a) (low intensity) and (b) (high intensity),  we find that the two distributions  are almost identical.

\begin{figure}
\begin{center}
\includegraphics[width=0.9\columnwidth]{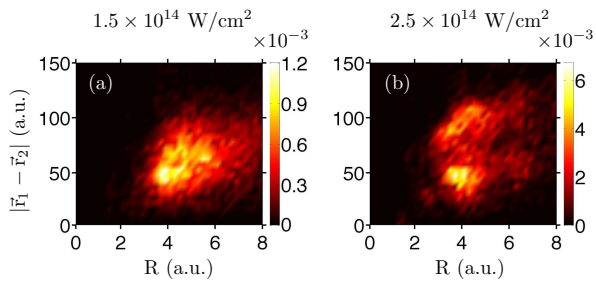}
\caption{(Color online) The 2-d distribution of the inter-electronic distance $\left\vert\vec{\mathrm{r}}_1-\vec{\mathrm{r}}_2\right\vert$ versus the inter-nuclear distance $\mathrm{R}$, at the time when the initially bound electron 2 tunnel-ionizes, for an intensity of $1.5\times10^{14}$ W$/$cm$^2$ (a) and for $2.5\times10^{14}$ W$/$cm$^2$ (b).}\label{fig3}
\label{doublediff}
\end{center}
\end{figure}

Next, we  ask how electron 2 gains energy to  transition from the ground state of the $\mathrm{H_{2}}$ molecule to a high Rydberg state of an H-atom. We find that electron 2  gains energy  through a strong interaction with the laser field that resembles ``frustrated" enhanced ionization in $\mathrm{H_{2}^{+}}$ (``frustrated" since electron 2 occupies a Rydberg state instead of escaping).  This resemblance is corroborated by  i) the final energy distribution being similar for double $\mathrm{H^{*}}$ formation [\fig{figure1}(c) and (d)]  and  enhanced ionization  \cite{Leth2009} and ii) by our finding that electron 2 preferentially tunnel-ionizes when the nuclei are roughly 5 a.u. apart as is the case for enhanced ionization. In \fig{doublediff} we plot the double differential probability of the inter-electronic distance versus the inter-nuclear distance at the time when electron 2 tunnel-ionizes. Indeed, we find that electron 2 preferentially  tunnel-ionizes when the nuclei are roughly 5 a.u. apart, with the inter-nuclear distance shifting to smaller values for the higher intensity.  This is similar to our finding in \cite{Emmanouilidou2012}, for single H$^{*}$ formation.

\begin{figure}
\begin{center}
\includegraphics[width=0.5\columnwidth]{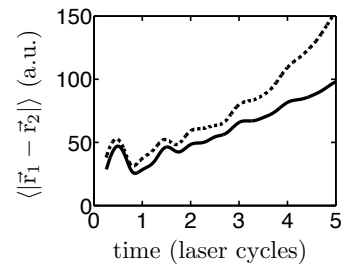}
\caption{The mean inter-electronic distance $\left\langle\left\vert\vec{\mathrm{r}}_1-\vec{\mathrm{r}}_2\right\vert\right\rangle$ as a function of time, for an intensity of $1.5\times10^{14}$ W$/$cm$^2$, for double H$^*$formation (solid), and for pathway B of single H$^{*}$ formation (dashed line).}\label{fig5}
\label{meand}
\end{center}
\end{figure}

The questions that naturally arise next are whether electronic correlations are present while two H$^{*}$ atoms form and whether these correlations are similar in strength to the electronic correlation in pathway B of single H$^{*}$ formation. In the latter case, we have shown 
that electron-electron correlation mostly resembles  the delayed pathway of NSDI \cite{Emmanouilidou2011NJP} (also referred to as re-collision-induced excitation with subsequent field ionization, RESI \cite{Kopold2000}) where the electronic correlation is weak.  For the delayed NSDI pathway, the re-colliding electron returns to the core close to a zero of the field, transfers energy to the second electron, and one electron escapes with a delay more than a quarter laser cycle after re-collision. Comparing, in \fig{meand}, the mean inter-electronic distance as a function of time for the pathway where two H$^{*}$ atoms form and for pathway B of single H$^{*}$ formation, we find that the two mean distances are very similar. Indeed, a soft re-collision is present in both cases. Moreover, the two electrons stay closer to each other during double H$^{*}$ formation rather than single H$^{*}$ formation; this is to be expected since in the former case it is both electrons that undergo ``frustrated" ionization oscillating in the vicinity of the nucleus. Thus, electron-electron correlation is similar in double H$^{*}$ formation and pathway B of single H$^{*}$ formation if not slightly stronger in the former case.

 Further, we find
that  the probability (out of all trajectories) for double H$^{*}$ formation  reduces from  
0.3\% for the lower intensity to 0.2\% for the higher one. Thus, forming two H$^{*}$ atoms is roughly 40-50 times more rare than forming one H$^{*}$ atom.  The reduction of the probability from 0.3\% to 0.2\% for double H$^{*}$ formation is consistent with a decrease, for increasing intensity, of electronic correlation  in the form of re-collisions. Indeed, when comparing \fig{doublediff}(a) with (b)
we find that at the time when electron 2 tunnel-ionizes the two electrons are much closer to each other (smaller inter-electronic distance) for the lower intensity.

\begin{figure}
\begin{center}
\includegraphics[width=0.9\columnwidth]{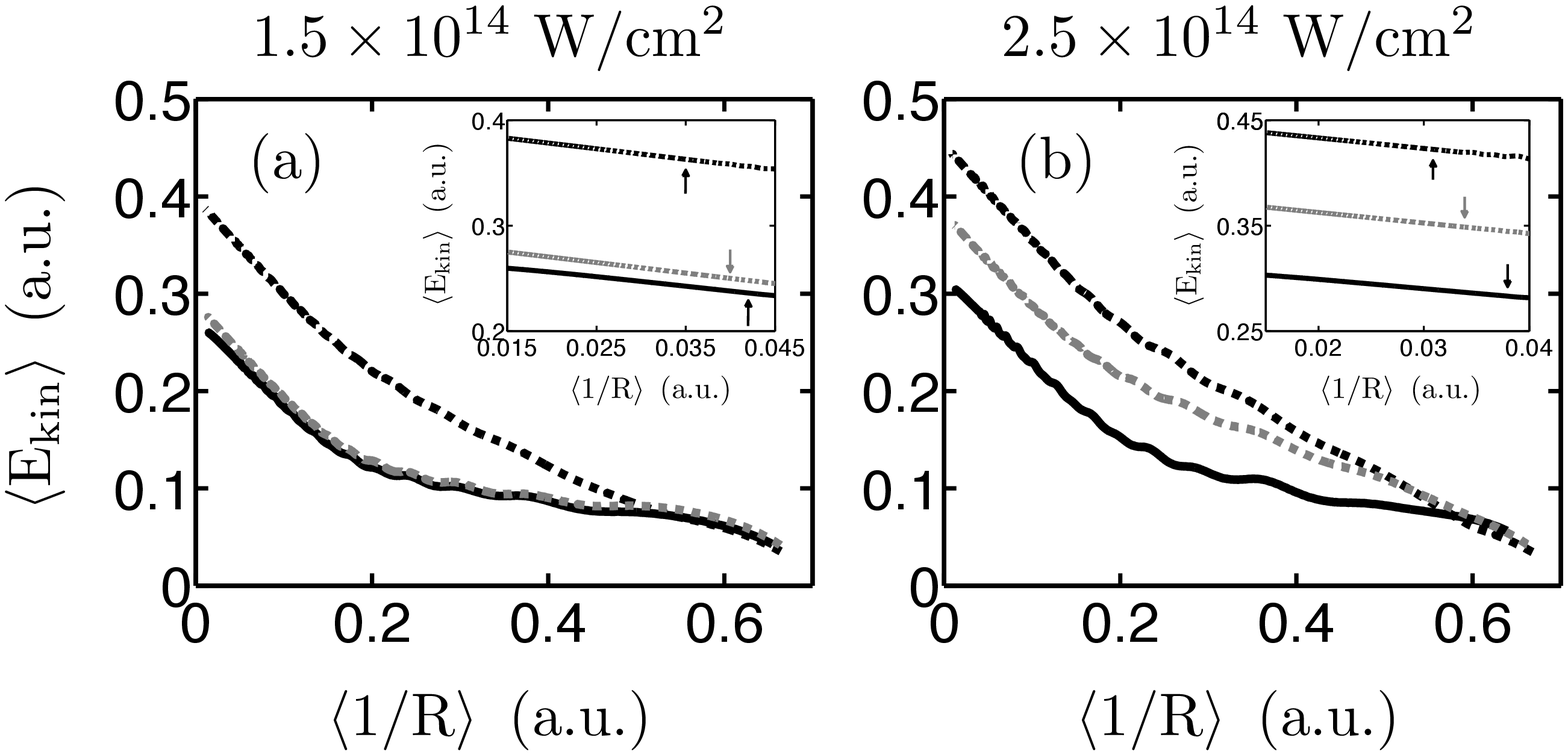}
\caption{The mean total kinetic energy of the nuclei $\left\langle \mathrm{E}_{\mathrm{kin}}\right\rangle$ versus the mean inverse  inter-nuclear distance $\left\langle1/\mathrm{R}\right\rangle$, for  $1.5\times10^{14}$ W$/$cm$^2$ (a) and for $2.5\times10^{14}$ W$/$cm$^2$ (b). The solid line is for the double H$^*$ formation channel, the dashed black/gray  line is for the H$^+$ fragments in the double ionization channel through re-scattering/enhanced ionization. The arrows indicate the $\left\langle1/\mathrm{R}\right\rangle$ corresponding to a time equal to 12T (duration of the laser pulse). The insets show the segments we fit for times larger than 12T to obtain slopes of 0.86  for the double H$^*$ formation channel and of 1 for the double ionization channels.  } 
\label{mean}
\end{center}
\end{figure}

We now gain further insight into the dynamics of the Coulomb exploding nuclei by plotting in \fig{mean} the mean total kinetic energy of the nuclei  $\langle \mathrm{E}_{\mathrm{kin}}\rangle$ versus the mean inverse inter-nuclear distance $\langle 1/\mathrm{R}\rangle$. Note that in \fig{mean} time increases as $\langle 1/\mathrm{R}\rangle$ decreases. From early on
up until the time (roughly 3.8 laser cycles) when electron 2 tunnel-ionizes at $\langle 1/\mathrm{R} \rangle \approx0.15$ a.u. it is mostly electron 2 that significantly screens the Coulomb repulsion of the two nuclei.  
 This screening is corroborated by the small slope of the  $\langle \mathrm{E}_{\mathrm{kin}}\rangle$ versus $\langle 1/\mathrm{R}\rangle$ curve for distances up to $\langle 1/\mathrm{R} \rangle \approx0.15$ a.u.
After electron 2 tunnel-ionizes the two nuclei move fast away from each other. Indeed, the slope  of  the $\langle \mathrm{E}_{\mathrm{kin}}\rangle$ versus $\langle 1/\mathrm{R}\rangle$ curve increases for  $\langle 1/\mathrm{R} \rangle > 0.15$ a.u to  0.86 for both intensities.

 This slope of 0.86 quantifies how much the two electrons that remain bound to highly excited states screen the Coulomb repulsion of the exploding nuclei at large inter-nuclear distances. 
To show that this is the case, we plot  $\langle \mathrm{E}_{\mathrm{kin}}\rangle$ versus $\langle 1/\mathrm{R}\rangle$ for break-up channels where both electrons escape. Such channels are the double ionization through re-collision or through enhanced ionization with two H$^{+}$ ions and two escaping electrons as the final fragments. To identify double ionization events through re-scattering, we use as a rough criterion the condition that electron 2 escapes without tunnel-ionizing. If, however, electron 2 tunnel-ionizes, we register these trajectories as double ionization through enhanced ionization events. \fig{mean} shows that the two nuclei move for all times faster away from each other in the double ionization channels rather than in the double H$^{*}$ formation channel. This is expected since in the former case  both electrons escape fast,  not screening
the Coulomb repulsion of the nuclei. Moreover, the nuclei move faster away from each other in the double electron escape through re-scattering rather than through the enhanced ionization channel. This is also expected since in the former case electron 2 escapes at times smaller than the time it takes for electron 2 to tunnel-ionize in the latter channel.  For large times (small $\langle 1/\mathrm{R}\rangle$), after the laser pulse is off, we find that the slope of  the $\langle \mathrm{E}_{\mathrm{kin}}\rangle$ versus $\langle 1/\mathrm{R}\rangle$  curve for the double ionization channels is 1, consistent with two H$^+$ nuclei Coulomb exploding unhindered by electronic motion.

Concluding, we have found that in double H$^{*}$ formation both tunnel-ionization steps are ``frustrated". We have shown that the route to forming two H$^{*}$ atoms is a pathway similar to pathway B of single H$^{*}$ formation. Specifically, two-electron effects are present in both single and double H$^{*}$ formation channels. That is, in double H$^{*}$ formation, electron 2 gains energy through a weak interaction with electron 1 resembling 
 delayed NSDI in H$_{2}$.  Moreover, electron 2 gains mostly energy  through a strong interaction with the laser field resembling ``frustrated" enhanced ionization of H$_{2}^{+}$, while electron 1 occupies a high Rydberg state. Two-electron effects diminish with increasing intensity with electron 2 gaining energy mainly through an interaction with the laser field. We emphasize that our 3-d quasiclassical formulation for describing break-up channels during Coulomb explosion of strongly driven H$_{2}$ is general and can be applied to study the break-up of driven multi-center molecules. 

\end{document}